\documentclass{elsart}
\usepackage{epsfig}
\begin{document}
\begin{frontmatter}
 \title{  Monitoring of the accelerator  beam distributions 
          for internal target facilities 
       }
\author[uj,ikp]{P.~Moskal\thanksref{email}},
\author[muenster]{H.-H.~Adam},
\author[ifj]{A.~Budzanowski},
\author[ikp]{D.~Grzonka},
\author[uj]{L.~Jarczyk},
\author[muenster]{A.~Khoukaz},
\author[ikp]{K.~Kilian},
\author[ikp,us]{P.~Kowina},
\author[muenster]{N.~Lang},
\author[muenster]{T.~Lister},
\author[ikp]{W.~Oelert},
\author[muenster]{C.~Quentmeier},
\author[muenster]{R.~Santo},
\author[ikp]{G.~Schepers},
\author[ikp]{T.~Sefzick},
\author[ikp]{S.~Sewerin\thanksref{swen}},
\author[us]{M.~Siemaszko},
\author[uj]{J.~Smyrski},
\author[uj]{A.~Strza{\l}kowski},
\author[ikp]{M.~Wolke},
\author[zel]{P.~W{\"u}stner},
\author[us]{W.~Zipper}
\thanks[email]{ Corresponding author. 
                {\em E-mail address}: p.moskal@fz-juelich.de
              }
\thanks[swen]{ present address: 
               The Svedberg Laboratory, S-75121 Uppsala, Sweden
             }
\address[uj]{ Institute of Physics, 
              Jagellonian University, PL-30-059 Cracow, Poland}
\address[ikp]{ IKP, Forschungszentrum J\"{u}lich, 
               D-52425 J\"{u}lich, Germany}
\address[muenster]{ IKP, Westf\"{a}lische Wilhelms--Universit\"{a}t,
                    D-48149 M\"unster, Germany}
\address[ifj]{ Institute of Nuclear Physics, PL-31-342 Cracow, Poland}
\address[us]{ Institute of Physics, University of Silesia, 
              PL-40-007 Katowice, Poland}
\address[zel]{ ZEL,  Forschungszentrum J\"{u}lich,
               D-52425 J\"{u}lich,  Germany}
\begin{abstract}
     We describe a direct method for monitoring the geometrical
     dimensions of a synchrotron beam at the target position
     for internal target installations.
     The method allows for the observation of the proton beam size
     as well as the position of the beam relative to the target.
     As a first demonstration of the technique,
     we present results obtained by means of the COSY-11
     detection system installed at the cooler synchrotron COSY.
     The influence of the stochastic cooling on the COSY proton beam
     dimensions is also investigated.

\end{abstract}
\end{frontmatter}
\section{Introduction}
\label{introduction}
Internal cluster target facilities -- such as COSY-11~\cite{brauksiepenim} installed
at the cooler synchrotron COSY-J\"ulich~\cite{maiernim}
-- permit the study of the production of mesons 
   in the proton-proton interaction
with high luminosity ($10^{~31}$~cm$^{-2}$ s$^{-1}$)
in spite of very low target 
densities ($\approx 10^{~14}$~atoms~cm$^{-2}$).
Such conditions minimize changes 
in the ejectiles' momentum vectors
due to secondary scattering in the target, and hence
facilitate the precise study  of reactions with cross section values
at the nanobarn level.

An exact extraction of absolute cross sections 
from the measured data demands
a reliable estimation of the acceptance of the detection system.
This in turn crucially depends on the accuracy of 
the determination of the position and
dimensions of both the beam and the target.
Using the COSY--11 detection 
system~\cite{brauksiepenim} as illustration,
we will describe a method for estimating the dimensions of
the  proton beam, based on the momentum distribution for
elastically scattered protons, which can be measured
simultaneously with the investigated reaction.

\begin{figure}[ht]
\leavevmode
\vspace{2cm}
\centerline{\epsfig{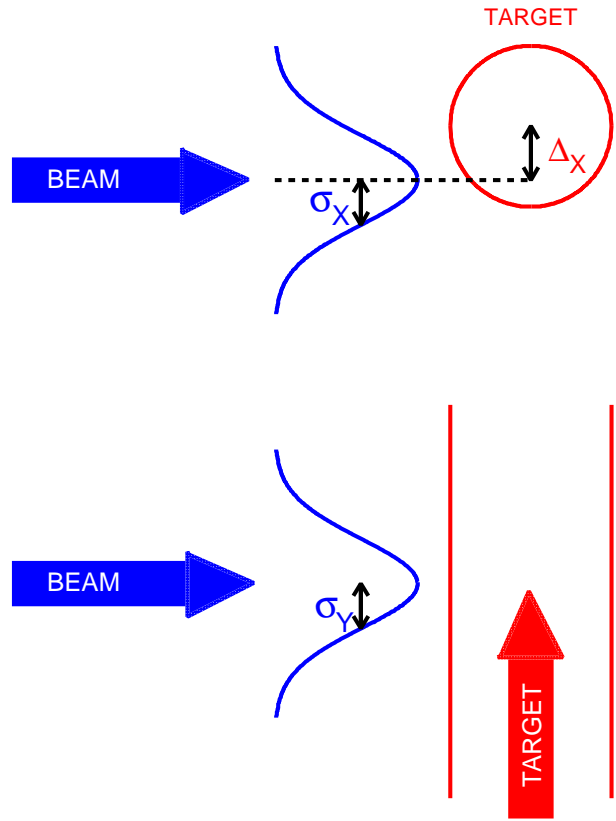}}
\vspace{1cm}
\caption{Schematic depiction of the relative beam and target
         setting. Seen from above (upper part), and from aside (lower
         part), $\sigma_{X}$ and $\sigma_{Y}$ 
         denote the horizontal and
         vertical standard deviation of 
         the assumed Gaussian distribution of
         the proton beam density, respectively. 
         The distance between the
         centres of the proton  and the target beam is described as
         $\Delta_{X}$.
        }
\label{kolo_gaussy}
\end{figure}

At the COSY--11 facility (see Fig.~\ref{elas_princip})
the production of short-lived uncharged mesons ($\eta$,
$\omega$, $\eta^{\prime}$, or $\phi$) 
and hyperons ($\Lambda$ and $\Sigma^{0}$)
is investigated by means of the
missing mass technique via the $pp\rightarrow ppX$ and
$pp\rightarrow pK^{+}Y$ reactions,
respectively~\cite{sewerinprl,smyrskipl,moskalpl}. In this way,
the four-momenta of positively charged particles
are fully determined experimentally.
The accuracy of the extracted missing mass value determines whether
the signal from the given meson is visible over a
background and depends on the precision of the momentum reconstruction of
the registered protons or kaons. Consequently, it varies with the detector
resolution, the momentum spread and the geometrical spread of the
accelerator proton beam interacting with the internal cluster target beam.
The momentum reconstruction is performed by tracing back trajectories
from  drift chambers through the dipole magnetic field to the
target~(Fig.~\ref{elas_princip},\cite{brauksiepenim}), 
which is an infinitely thin vertical line in the ideal case.
In reality, however, the reactions take place in a region of finite
dimensions where the beam and the target overlap~(see Fig.~\ref{kolo_gaussy}).
Assuming an infinitesimal target, our analysis shows a smearing out of the 
momentum vectors and hence of the resolution of the missing mass signal.
However, if the target and beam dimensions are known it is
possible to determine the average smearing of the missing mass
originating from this effect and hence to infer 
for example the natural width of the measured mesons,
provided that the beam momentum spread is known as well.
\section{Description of the method}
\label{description}
\begin{figure}[ht]
\leavevmode
\vspace{2cm}
\centerline{\epsfig{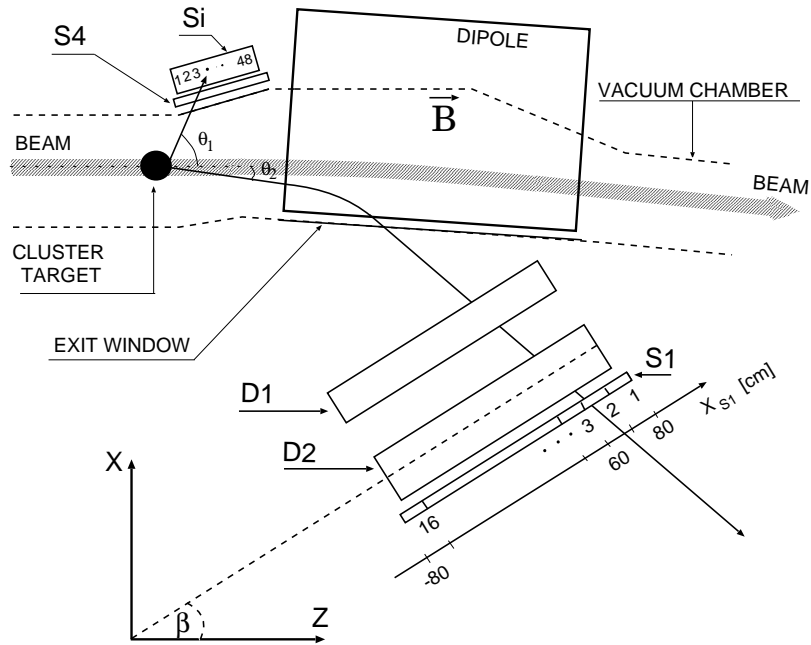}}
\vspace{1cm}
\caption{Schematic diagram of the COSY--11 detection setup. Only
 detectors used for the measurement of elastically scattered protons
 are shown. Numbers used on the silicon pad detector (Si) and below the
 scintillator hodoscope (S1) indicate the order of segments. D1 and
 D2 denote drift chambers. The X$_{S1}$  axis is defined such that
 the first segment of the S1 ends at 80~cm and the sixteenth ends at
 -80~cm. The proton beam, 
 depicted by a shaded line, circulates in the 
 ring and crosses each time the $H_{2}$ 
 cluster target installed in front
 of one of the bending dipole magnets of the COSY accelerator.
 }
\label{elas_princip}
\end{figure}

\begin{figure}[hbt]
\leavevmode
\vspace{2cm}
\centerline{\epsfig{file=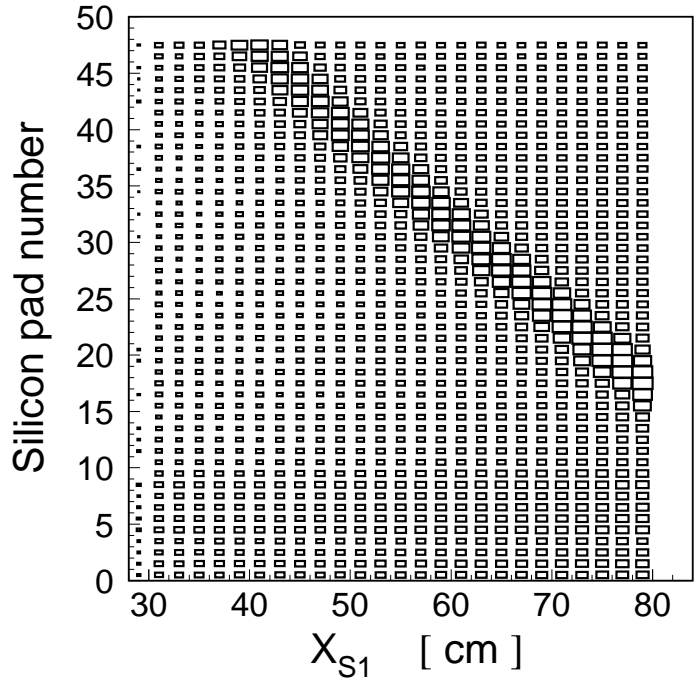,width=11cm}}
\vspace{1cm}
\caption{ Identification of elastically scattered protons 
          from the correlation
          of hits in the silicon detector Si 
          and the S1 scintillator hodoscope.
          Note that the number of entries per bin is given in
          a logarithmic scale, 
          ranging from 1 (smallest box) to 19000 (largest box).
        }
\label{correl}
\end{figure}

\begin{figure}[hbt]
 \leavevmode
 \unitlength 1.0cm 
  \begin{picture}(12.2,14.0)
    \put(2.5,0.0){
       \epsfig{figure=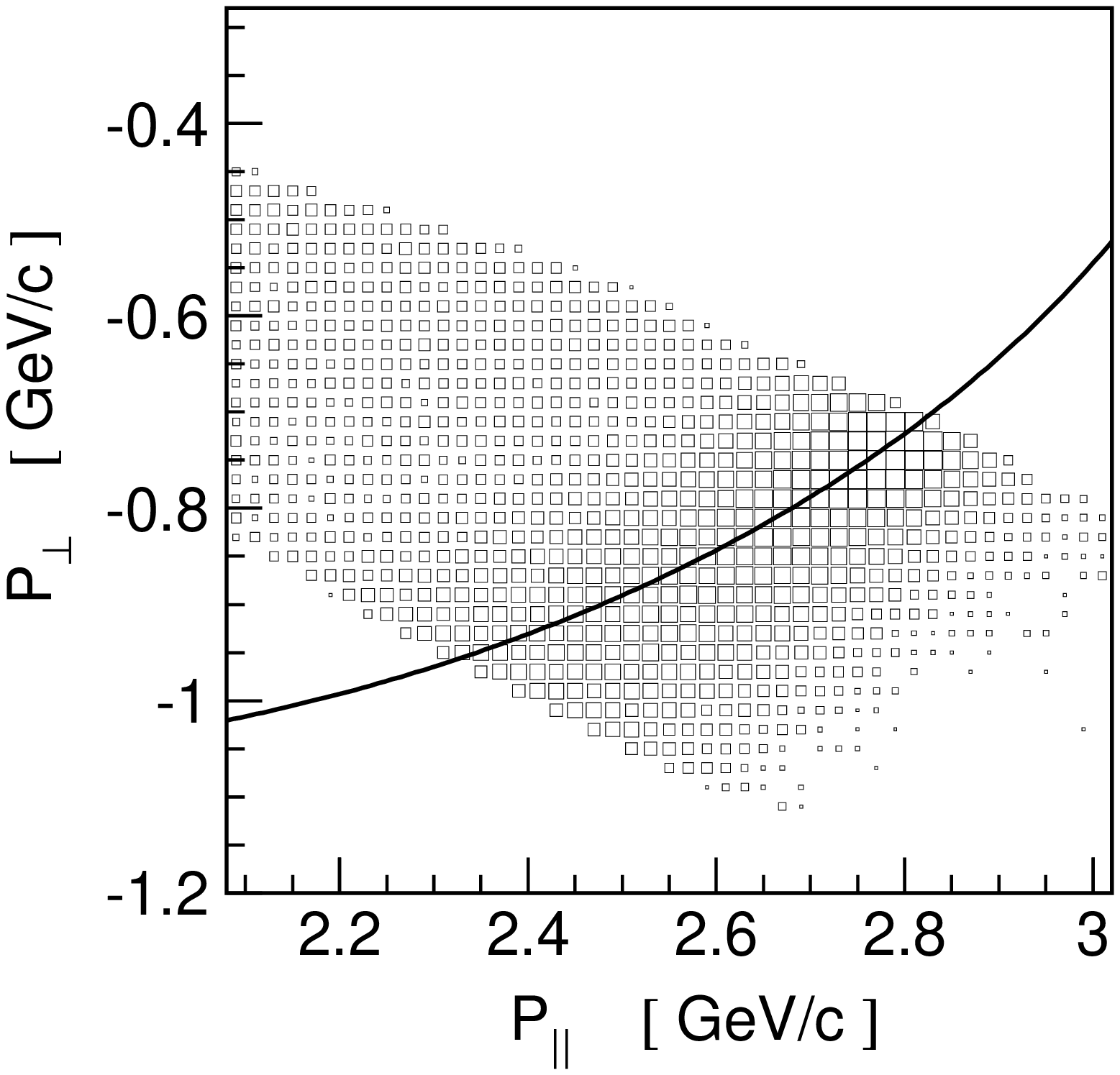,height=9.0cm,angle=0}
    }
       \put(10.3,1.5){
          { b)}
       }
    \put(2.5,7.0){
       \epsfig{figure=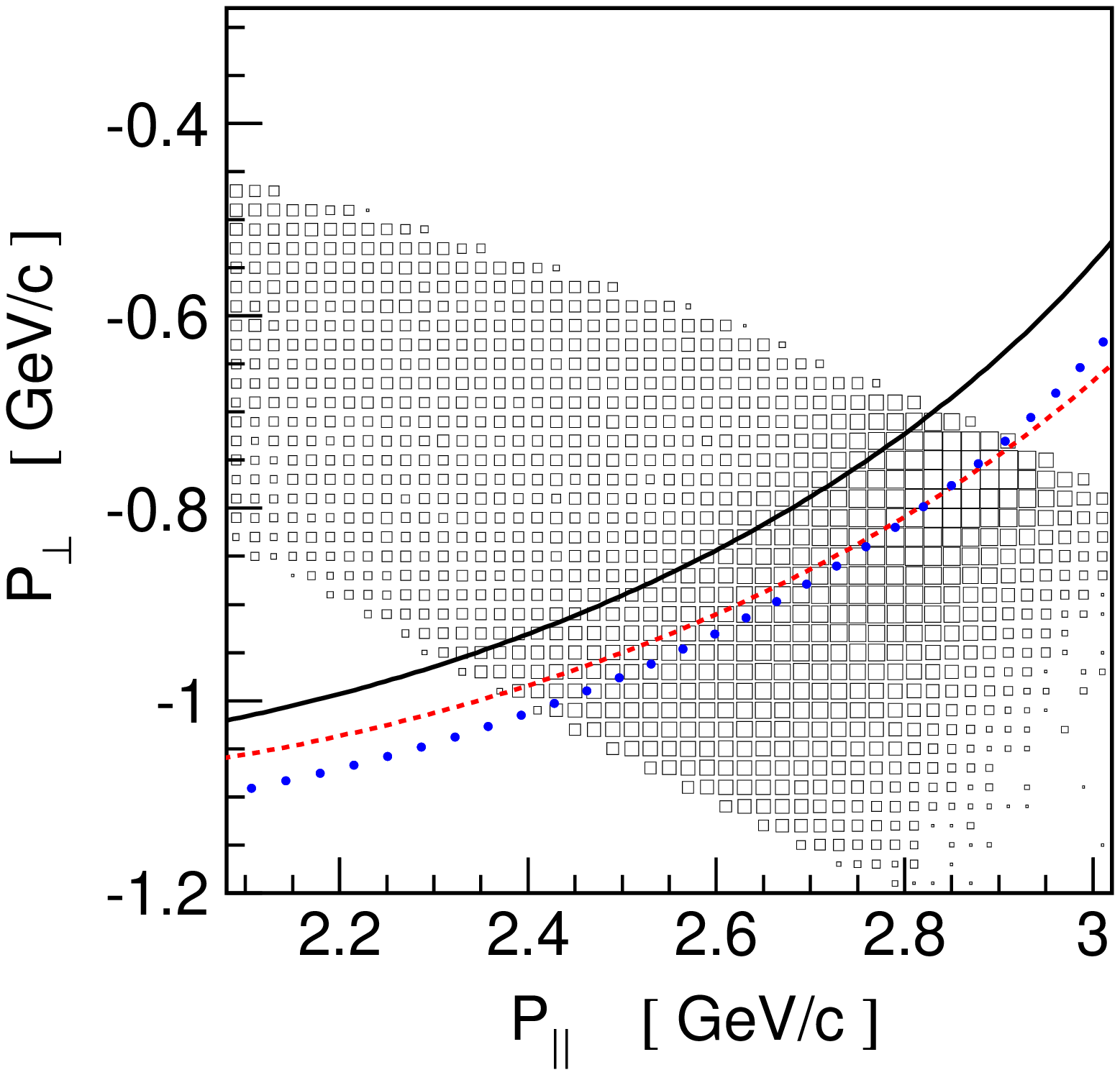,height=9.0cm,angle=0}
    }
       \put(10.3,8.5){
          { a)}
       }
  \end{picture}
\caption{  a) \ \ Perpendicular versus parallel 
               momentum components with
               respect to the beam direction 
               of particles registered at a beam
               momentum of 3.227~GeV/c. 
               The number of entries per bin is shown
               logarithmically. 
               The solid line corresponds to the momentum ellipse
               expected for elastically scattered protons 
               at a beam momentum of 3.227~GeV/c,
               the dashed line refers to a beam momentum of
               3.350~GeV/c, and the dotted line 
               shows the momentum ellipse
               obtained for a proton beam 
               inclined by 40~mrad. \protect\\
           b ) \ \ \ The same data as shown in a) but analyzed
               with the target point shifted by -0.2~cm  
               perpendicularly to the beam
               direction (along the X-axis in  
               Figure~\protect\ref{elas_princip}). 
               The solid line shows the momentum 
               ellipse at a beam momentum of 3.227~GeV/c.
        }
 \label{trans_paral}
\end{figure}

Part of the COSY--11 detection setup used for the registration of
elastically scattered protons is shown in Figure~\ref{elas_princip}.
Trajectories of protons scattered in the forward direction are
measured by means of two drift chambers (D1 and D2)
and a scintillator hodoscope~(S1),
whereas the recoil protons are registered in coincidence with the
forward ones using a silicon pad detector arrangement (Si) 
and a scintillation detector~(S4).
The two--body kinematics gives an unambiguous relation between the
scattering angles $\Theta_{1}$ and $\Theta_{2}$
of the recoil and forward
protons (see Fig.~\ref{elas_princip}).
Therefore, as seen in Figure~\ref{correl}, events of
elastically scattered protons 
can be identified from the correlation
line formed between the position in the silicon pad detector Si, 
and the scintillator hodoscope S1,
the latter measured by the two drift chamber stacks.
For those protons which are elastically scattered 
in the forward direction and  deflected
in the magnetic field of the dipole, 
the momentum vector at the target point
can be determined.
According to two--body kinematics, momentum components parallel and
perpendicular to the beam axis form an ellipse. A section of this ellipse
is shown as a solid line in Figure~\ref{trans_paral}a
superimposed on the data which were selected according to the correlation
criterion from Figure~\ref{correl} for elastically scattered events.

In Figures~\ref{correl} and Figure~\ref{trans_paral}a, 
the elastically scattered events stand out clearly against a
low level background. But more importantly the mean of the
elastically scattered data is significantly 
shifted from the expected line, indicating that the
reconstructed momenta are on average larger than expected.
This discrepancy is difficult to explain in terms of the uncertainties 
of either the proton beam momentum or the proton beam angle. 
For instance, to account for this discrepancy, 
the beam momentum must be changed by more than 120~MeV/c 
(dashed line in Figure~\ref{trans_paral}a),
which is 40~times larger than the conservatively estimated error of the
absolute beam momentum~($\pm$~3~MeV/c)~\cite{prasuhn,moskalann}.
Similarly, the effect could have been corrected by changing the
beam angle by 40~mrad~(see dotted line in Figure~\ref{trans_paral}a),
which exceeds the uncertainty of the beam
angle~($\pm$~1~mrad~\cite{prasuhn}) by at least a factor of~40.

A reasonable agreement is obtained, however,  by shifting the
assumed reaction point relative to the nominal value by $-0.2$~cm
perpendicular to the beam axis towards the center of the COSY--ring, along
the X-axis defined in Figure~\ref{elas_princip}.
The experimentally extracted momentum components obtained in this manner
are shown in Figure~\ref{trans_paral}b, together with the expectation 
depicted by the solid line. There is good agreement and the data are now 
spread symmetrically around the ellipse.
   This spread is essentially due to
   the finite extensions of the cluster 
   target and the proton beam overlap
   (about $\pm 0.2$~cm as inferred from the shift in
   the target position required).
Other contributions, including the spread of the beam momentum 
and still some multiple scattering events, appear to be negligible.
We now consider in more detail the influence of the effective target 
dimensions and the spread of the COSY beam.

We assume  the target beam to be described by a cylindrical pipe,
with a diameter of 9~mm, homogeneously filled with 
protons~\cite{dombrowskinim,khoukazepj}. Furthermore,
the COSY proton beam density distribution is described by Gaussian 
functions with standard deviations
$\sigma_{X}$ and $\sigma_{Y}$ for the horizontal and vertical
directions (see Fig.~\ref{kolo_gaussy}), 
respectively~\cite{prasuhnnim,schulzphd}.
Monte--Carlo calculations are performed varying the distance between
the target  and the beam centres ($\Delta_{X}$), and the
horizontal proton beam extension ($\sigma_{X}$).

In order to account for the angular distribution 
of the elastically scattered protons, an appropriate weight $w$ 
was assigned to each generated event according
to the  differential distributions of the cross sections
measured by the EDDA collaboration~\cite{albersprl}.
The generated events were  analyzed in the same way as the
experimental data.
A comparison of the data from 
Figure~\ref{trans_paral}a (data analyzed 
using the nominal interaction point) with the
corresponding simulated histograms allows both
$\sigma_{X}$  and $\Delta_{X}$ to be determined.
 For finding an estimate of the parameters 
$\sigma_{X}$ and $\Delta_{X}$
we construct the $\chi^{2}$ statistic 
according to the {\em method of least squares}:
\begin{equation}
\chi^2 =  \sum_i \, \frac{
                          \left( \alpha N_i^s + b_i - N_i^e 
                          \right)^{2} 
                         } 
                         { \alpha^{2} \displaystyle 
                           \sum_{i^{th}bin} w^2 + N_i^e + b_i
                         },
\label{eqchi2_sl}
\end{equation}
where  $N_i^e$ and
${\displaystyle N_i^s = \sum_{i^{th}bin} w }$
denote the content of the $i^{th}$ bin of
the P$_{\perp}$--versus--P$_{\parallel}$ spectrum determined from
experiment (Fig.~\ref{trans_paral}a) and simulations, respectively. 
The background events $b_i$ in the i$^{th}$ bin amount 
to less than one per cent of the
data~\cite{moskalphd}
and were estimated by linear interpolations 
between the inner and outer part
of a broad distribution surrounding the expected 
ellipse originating from elastically
scattered protons. 
The free parameter $\alpha$ allows adjustment of the overall scale of
the fitted Monte--Carlo histograms. 
Thus, by varying the $\alpha$ parameter, the
$\chi^{2}_{min}$
for each pair of $\sigma_{X}$ and $\Delta_{X}$ was established
as a minimum of the $\chi^{2}(\alpha)$ distribution.

\begin{figure}[hbt]
 \leavevmode
 \unitlength 1.0cm 
  \begin{picture}(12.2,16.0)
    \put(8.0,10.0){
       \epsfig{figure=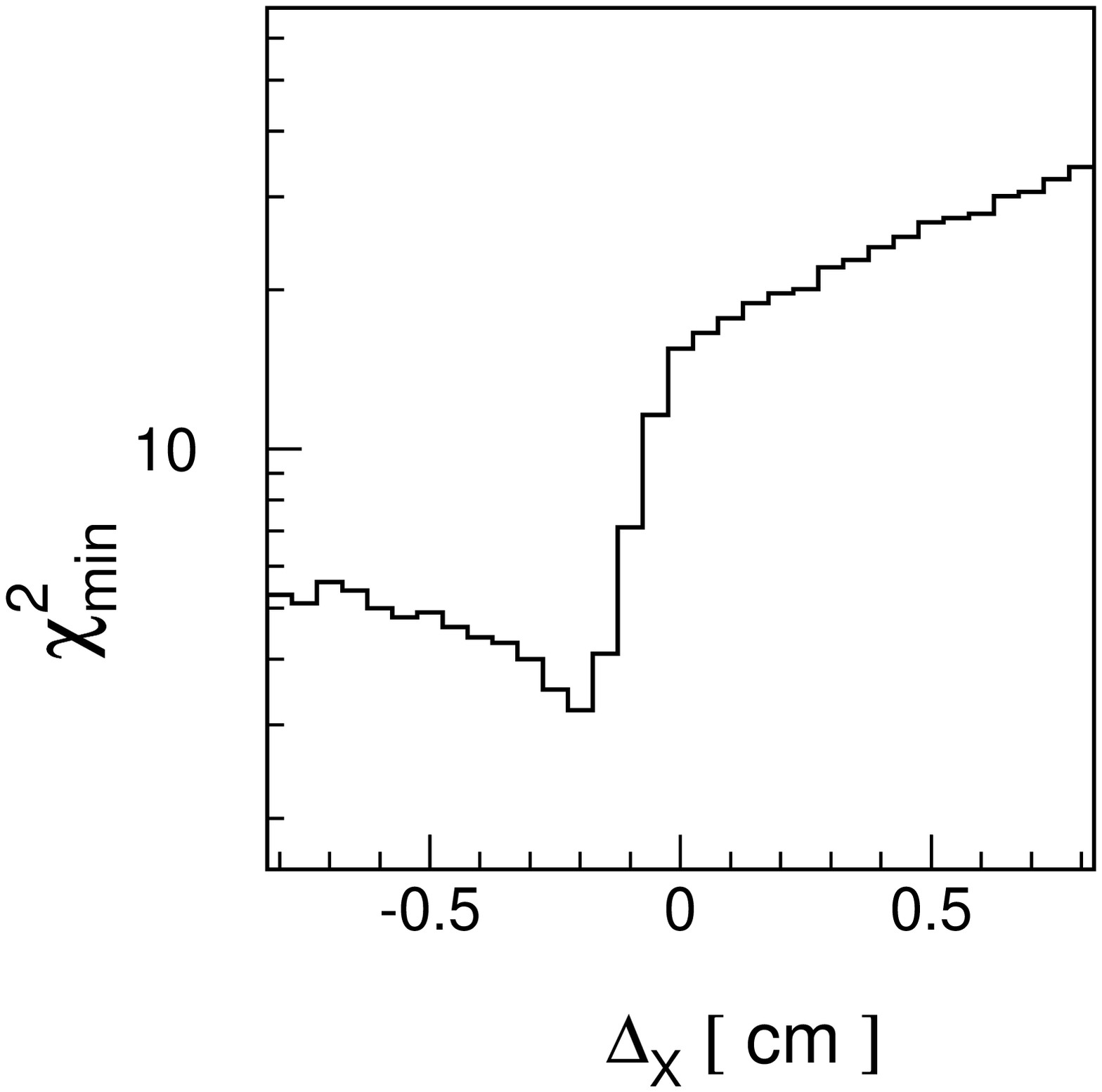,height=8.0cm,angle=0}
    }
       \put(15.,11.5){
          { b)}
       }
    \put(-0.5,3.5){
       \epsfig{figure=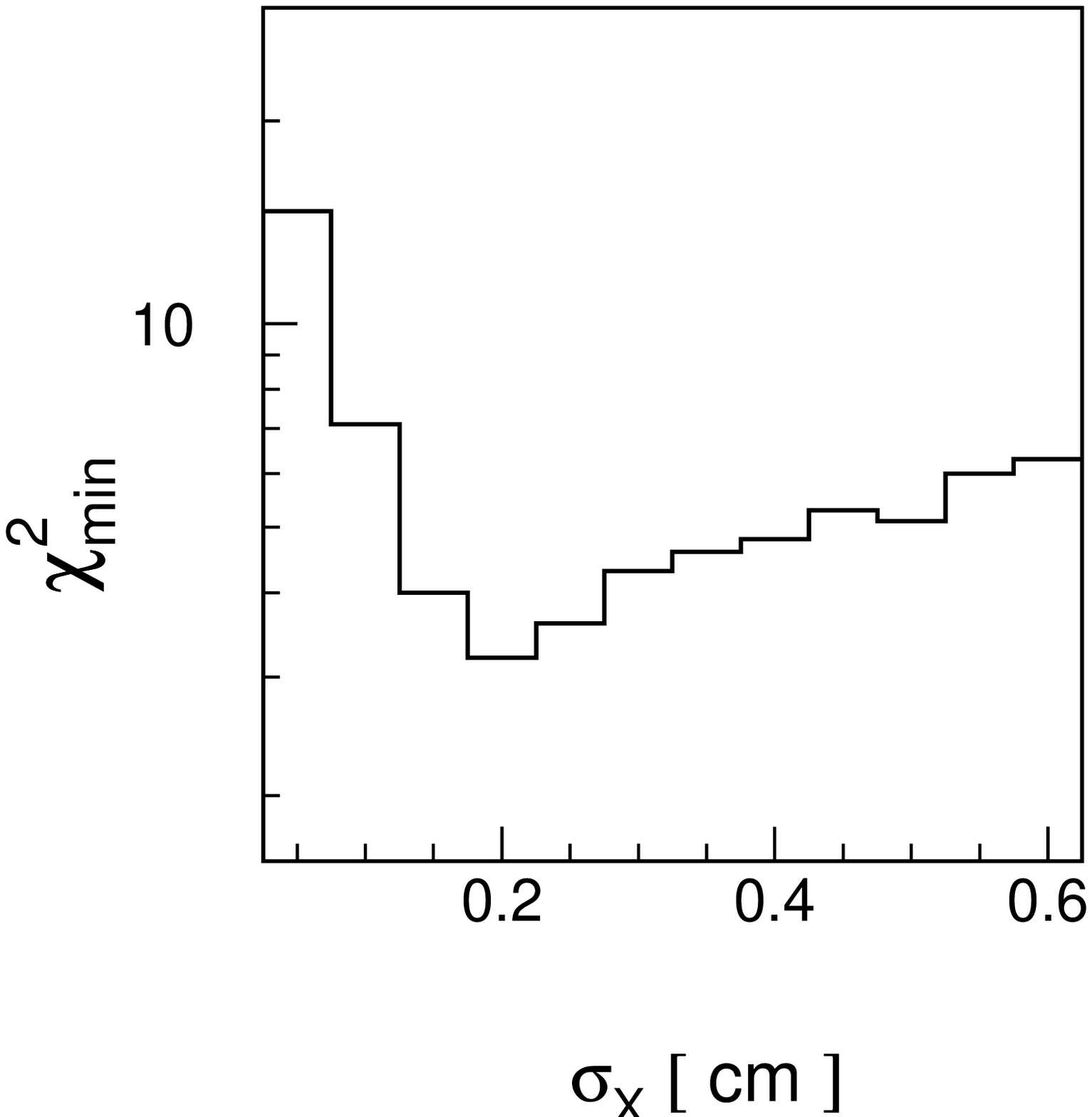,height=8.0cm,angle=0}
    }
       \put(6.5,5.0){
          { c)}
       }
    \put(-0.5,10.0){
       \epsfig{figure=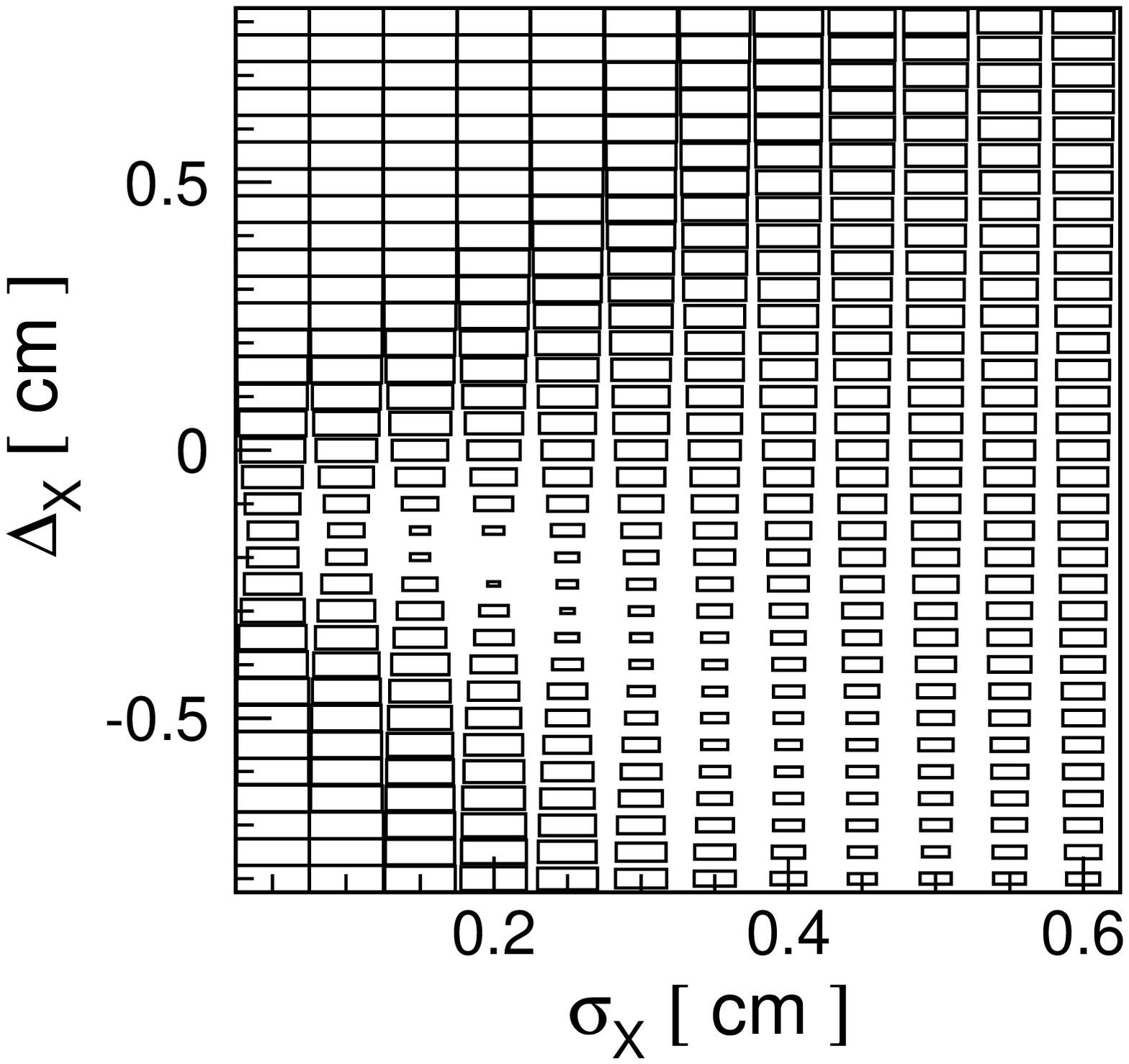,height=8.0cm,angle=0}
    }
       \put(6.5,11.5){
          { a)}
       }
  \end{picture}
 \caption{ 
          a) $\chi^{2}_{min}$ as a function 
            of $\sigma_{X}$ and $\Delta_{X}$. 
            The number of entries is shown 
            in  a logarithmic scale.  \ \ 
          b) $\chi^{2}_{min}$ as a function of $\Delta_{X}$. \ \ \
          c) $\chi^{2}_{min}$ as a function of $\sigma_{X}$. \
         The minimum value of   
         $\chi^{2}_{min}$ is larger than two. 
         However, it decreases when only a part 
         of the 60 minutes long COSY cycle
         is taken into account (see section~\ref{monitoring}).
         This is due to the fact that the beam changes
         during the cycle and  its form, 
         when integrated over the whole cycle time,
         does not suit perfectly to the form 
         assumed in Monte--Carlo simulations.
        }
 \label{dx1_vs_sx1}
\end{figure}

The results of the Monte--Carlo calculations are shown in
Figure~\ref{dx1_vs_sx1}.\\
Figure~\ref{dx1_vs_sx1}a shows the logarithm 
of $\chi^{2}_{min}$ as a function of the values of 
$\sigma_{X}$ and $\Delta_{X}$.  
The $\chi^{2}_{min}$ function has a valley identifying a minimum and allowing
the unique determination of the 
varied parameters~$\sigma_{X}$ and $\Delta_{X}$.
The overall minimum of the
$\chi^{2}_{min}(\sigma_{X},\Delta_{X})$--distribution, at
$\sigma_{X}$~=~0.2~cm and $\Delta_{X}$~=~-~0.2~cm is more evident in
Figures~\ref{dx1_vs_sx1}b~and~\ref{dx1_vs_sx1}c, 
which show the projections of
the valley line onto the respective axis.

The same results were obtained when employing
 the Poisson likelihood $\chi^{2}$ derived from the
maximum likelihood method~\cite{bakernim,feldmanpr}
\begin{equation}
\chi^2 = 2 \cdot \sum_i \, [\alpha N_i^s + b_i - N_i^e +  N_i^e \,
                            ln(\frac{N_i^e}{\alpha N_i^s + b_i})],
\label{eqchi2_mh}
\end{equation}

\begin{figure}[hbt]
\leavevmode
\centerline{\epsfig{file=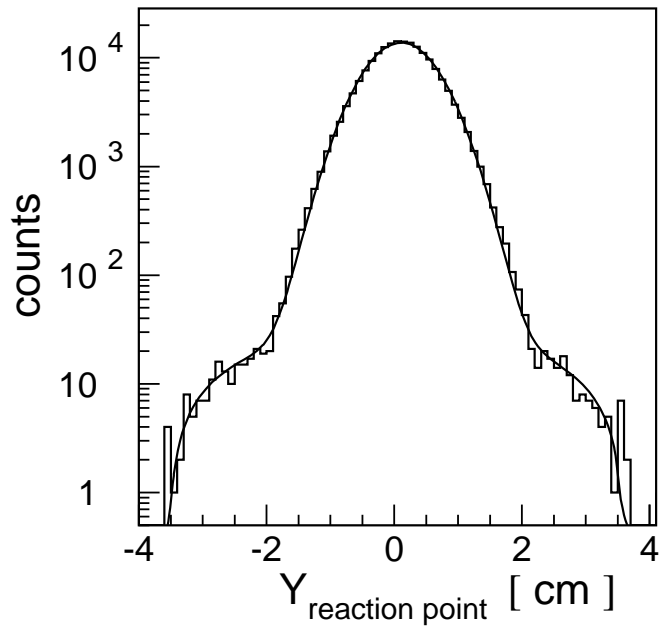,height=11cm,width=11cm}}
\vspace{1cm}
\caption{Distribution of the vertical component of the reaction
points determined by tracing back trajectories from the drift
chambers through the dipole magnetic field to the centre of the
target (in the horizontal plane). ``Tails'' are due to secondary
scattering on the vacuum chamber and were parametrized by a
polynomial of second order. The solid line shows the simultaneous fit
of the Gaussian distribution and the polynomial of second order.}
\label{y_tar}
\end{figure}

The vertical beam extension of $\sigma_{Y}$~=~0.51~cm was established
directly from the distribution of the vertical component of the
particle trajectories at the centre of the target.
This is possible, since for the momentum reconstruction only the
origin of the track in the horizontal plane is used, while the
vertical component remains a free parameter.
As shown in Figure~\ref{y_tar}, the reconstructed distribution
of the vertical component of the reaction points indeed can be
described well by a Gaussian distribution with the $\sigma$~=~0.53~cm.
 The width of this distribution
is primarily due to the  vertical spread of a proton beam.
The spread caused by
multiple scattering and  drift chambers resolution 
was determined to be
about 0.13~cm (standard deviation)~\cite{moskalphd}.
Therefore, $\sigma_{Y}~=~\displaystyle\sqrt{\displaystyle 0.53^{2} -
0.13^{2}}~\approx$~0.51~cm.

The effect of a possible drift chamber misalignment, i.\ e.\ an
inexactness of the angle $\beta$ in Figure~\ref{elas_princip}, was
estimated to cause a shift in the momentum plane by 
0.15~cm for $\Delta_{X}$.
This gives a rather large systematical bias in the estimation of the
absolute value of $\Delta_{X}$, but it does not affect
the parameter~$\sigma_{X}$, and still allows
for the determination of relative beam shifts $\Delta_{X}$ during
the measurement cycle.
\section{Monitoring of the beam geometry during the measurement cycle}
\label{monitoring}
\begin{figure}[hbt]
 \leavevmode
 \unitlength 1.0cm
  \begin{picture}(12.2,14.0)
    \put(-0.5,0.0){
       \epsfig{figure=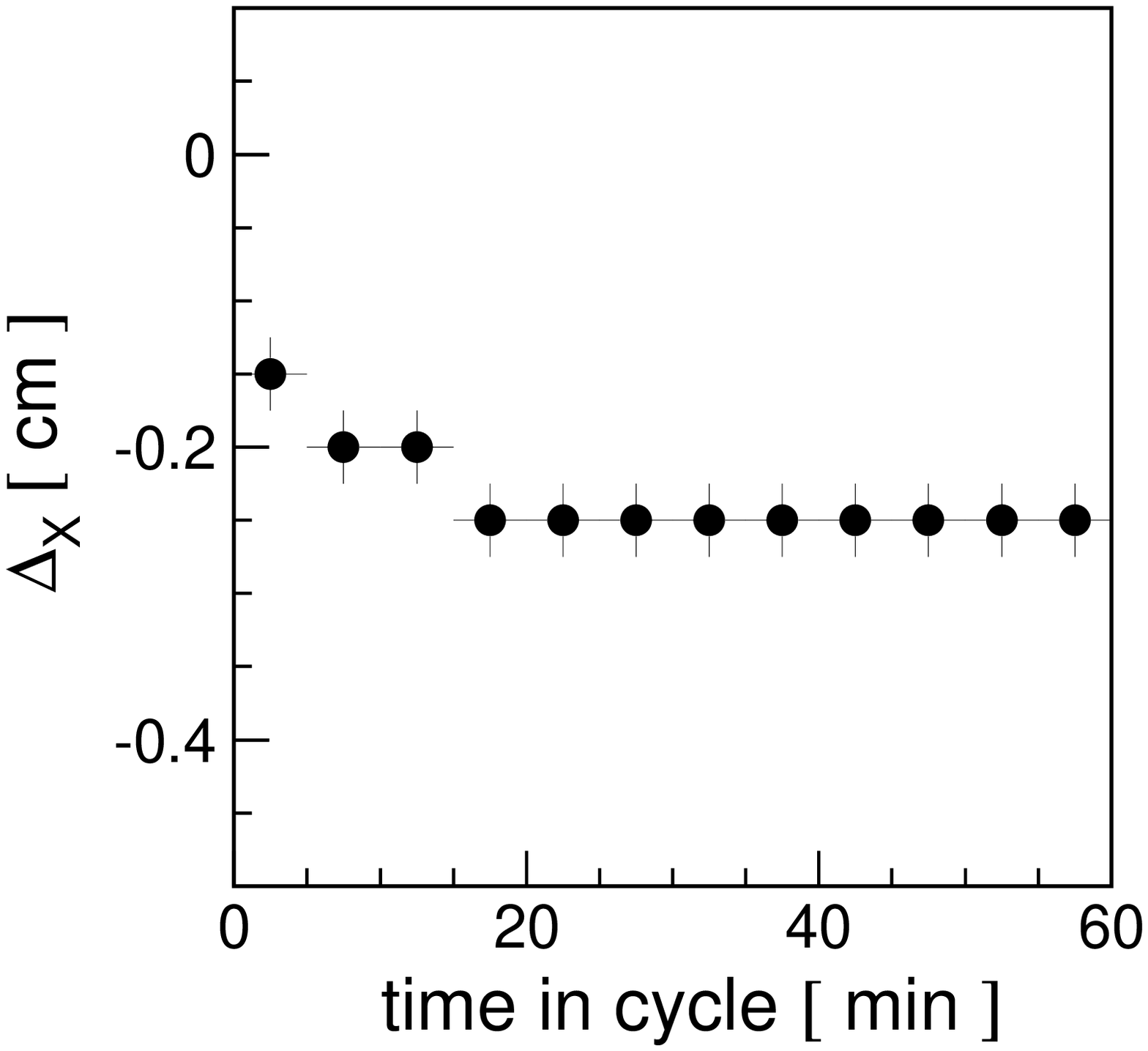,height=9.0cm,angle=0}
    }
       \put(7.2,1.7){
          { c)}
       }
    \put(-0.5,7.0){
       \epsfig{figure=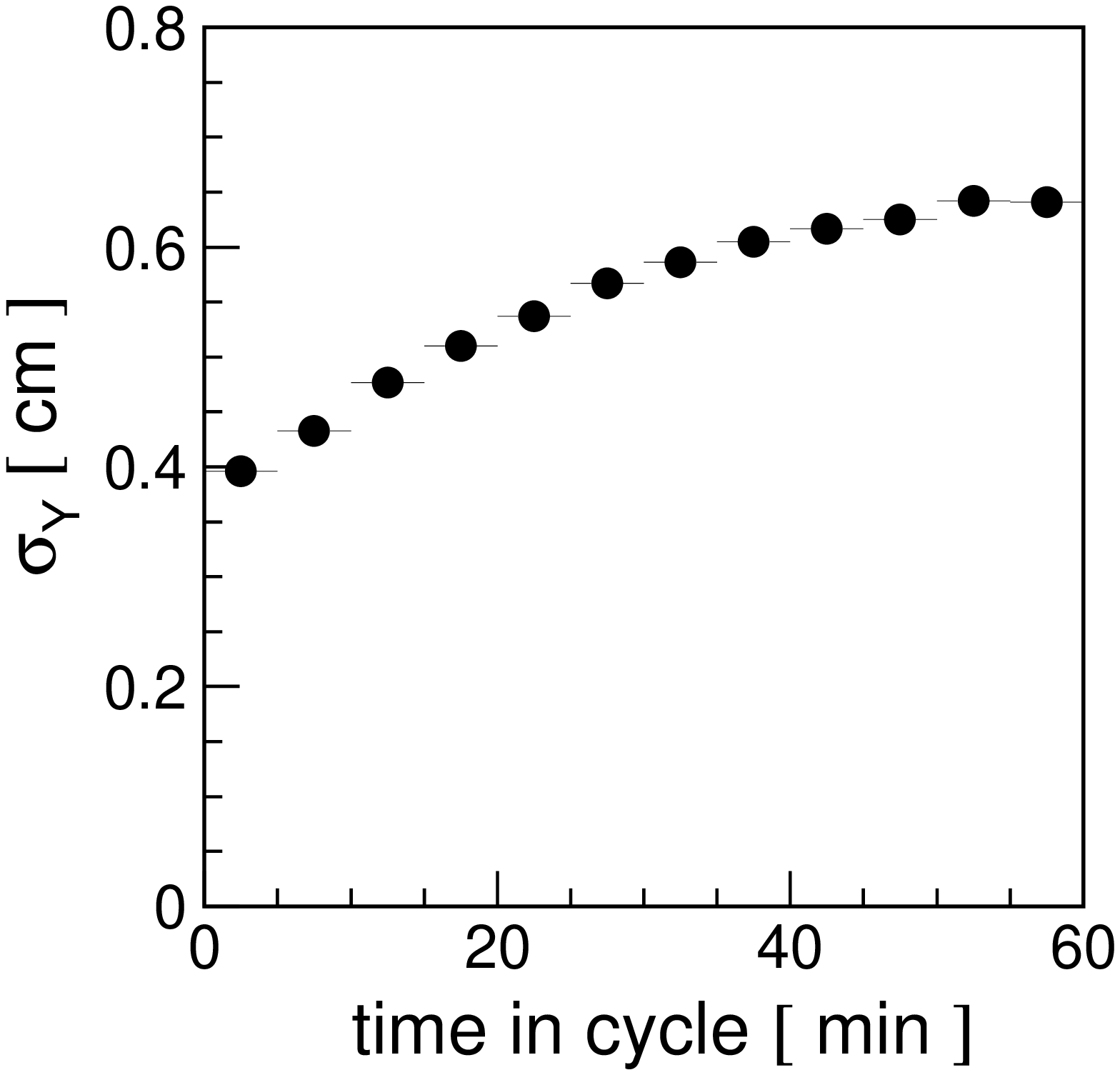,height=9.0cm,angle=0}
    }
       \put(7.2,8.7){
          { a)}
       }
    \put(8.0,0.0){
       \epsfig{figure=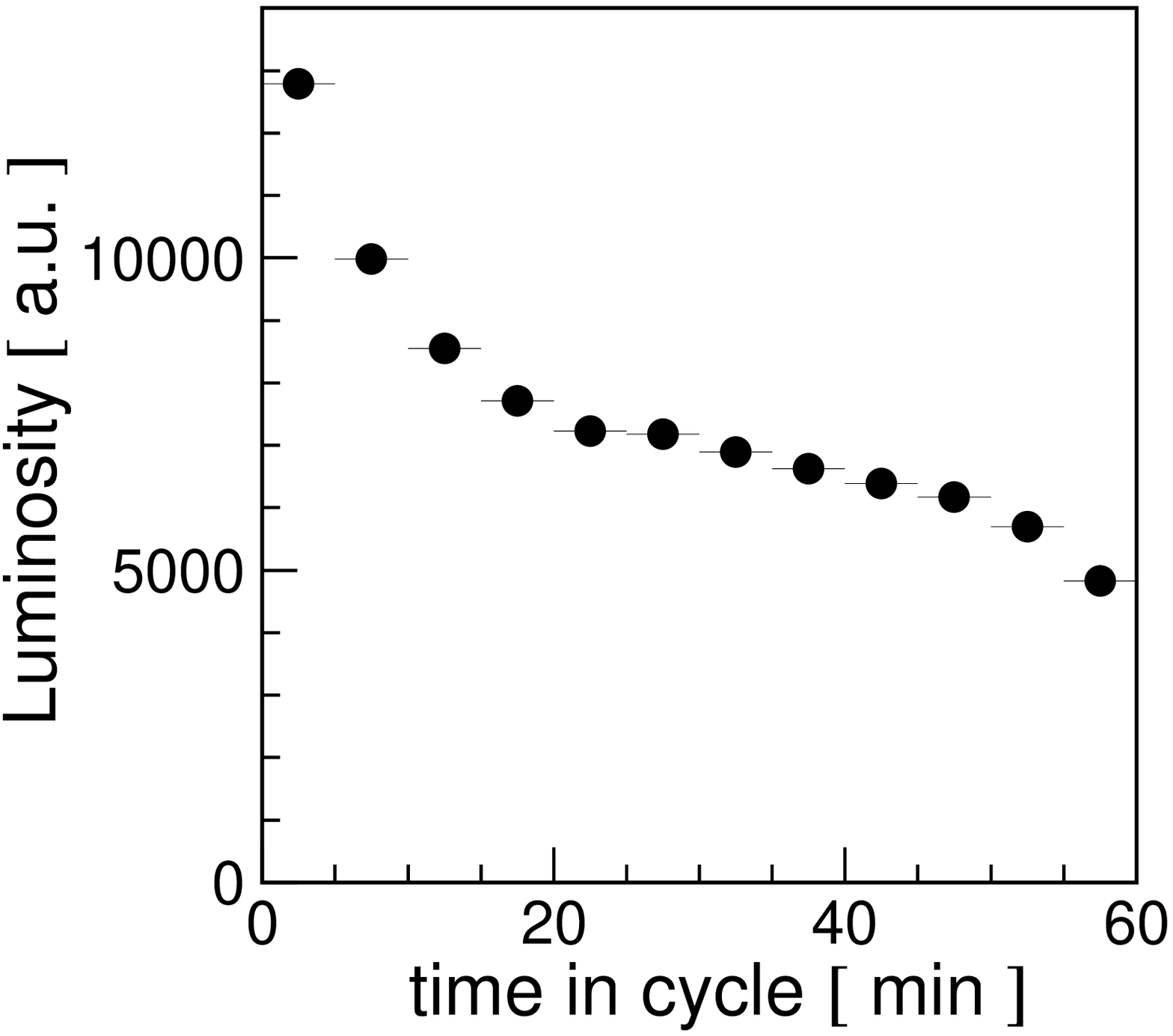,height=9.0cm,angle=0}
    }
       \put(15.7,1.7){
          { d)}
       }
    \put(8.0,7.0){
       \epsfig{figure=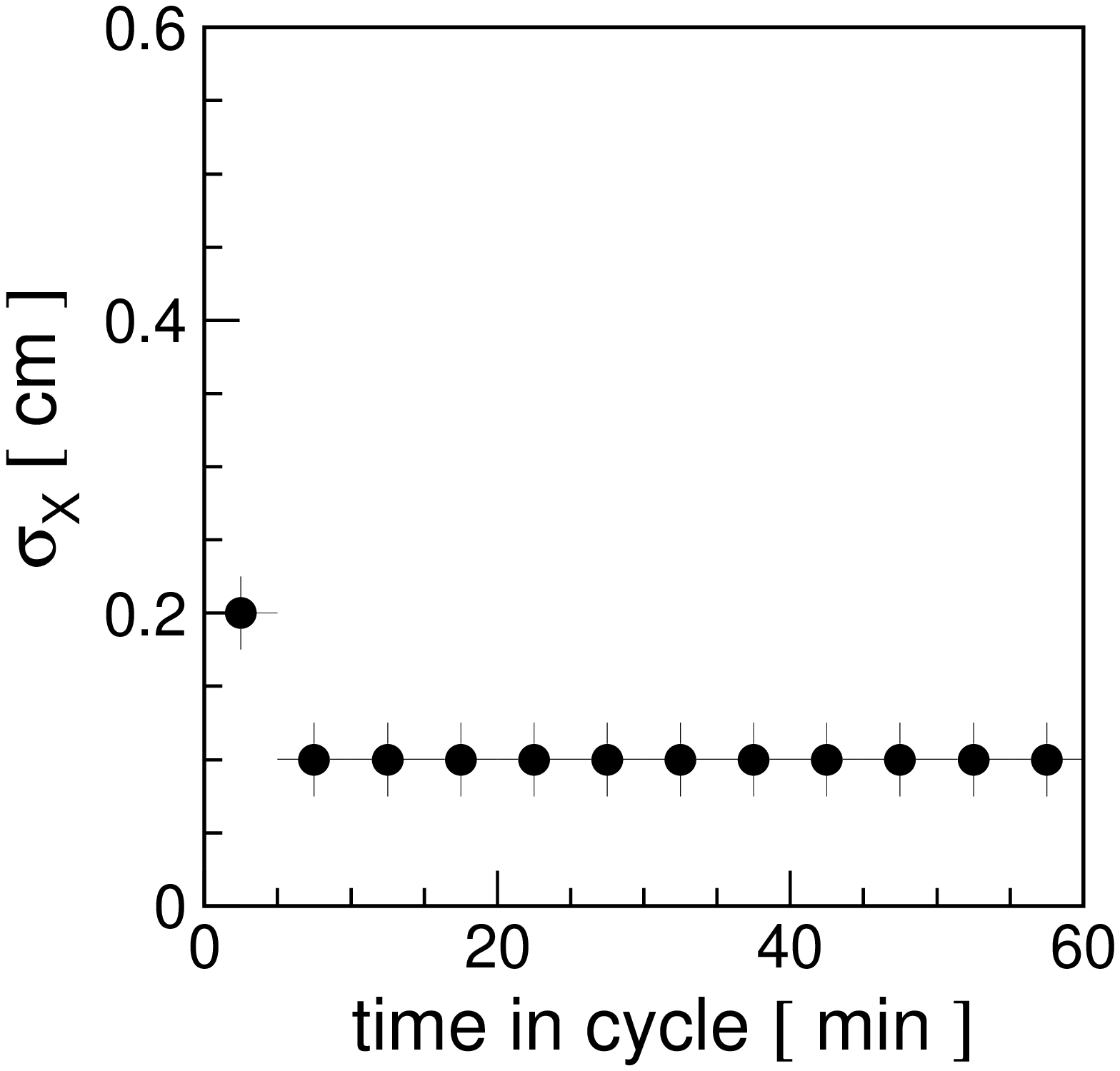,height=9.0cm,angle=0}
    }
       \put(15.7,8.7){
          { b)}
       }
  \end{picture}
\caption{ The vertical a) and horizontal b) 
          beam width (one standard deviation)
          determined for  each five minutes 
          partition of the COSY cycle. \protect\\
          Note that the beam width $\sigma_{X}$ 
          determined for each five minutes period is smaller than the 
          $\sigma_{X}$  averaged over the whole cycle 
          (compare Figure~\ref{dx1_vs_sx1}c).
          This is due to the beam shift relative to the target during 
          the experimental cycle as shown in the panel 
          on the bottom left. \protect\\
          c) Relative settings of the  COSY proton beam 
          and the target centre versus 
          the time of the measurement cycle.
          d) Changes of the luminosity during the cycle.
          A steep decrease on the beginning is 
          caused by the movement of the beam  
          out from the target centre.\protect\\
          The vertical error bars in pictures b) and c)  
          denote the size of the 
          step used in the Monte-Carlo simulations 
          ($\pm$~0.025~cm ; the bin width of 
           Figures~\ref{dx1_vs_sx1}b and~\ref{dx1_vs_sx1}c).
        }
 \label{sx_sy_cykl}
\end{figure}

It is possible to use the method described here to control the size and 
position of the beam relative to the target. This was demonstrated 
during the experimental run performed in February 1998~\cite{moskalpl},
when for the first time longitudinal and horizontal
stochastic cooling were used at the COSY accelerator. \\
The horizontal stochastic cooling~\cite{prasuhnnim} is used to squeeze
the proton beam in the horizontal direction until the beam reaches the
equilibrium between the cooling and the heating due to the target.
The size of an uncooled beam increases during the cycle,
as can be seen in Figure~\ref{sx_sy_cykl}a, which shows the
spreading of the beam in the vertical plane during the 60 minutes cycle.
This was expected since stochastic cooling in the vertical plane was not 
used in this case.
The influence of the applied cooling in the horizontal
plane is clearly visible in Figure~\ref{sx_sy_cykl}b.
During the first five minutes of the cycle the horizontal size of the beam 
of about  $2\cdot 10^{10}$ protons  was reduced by a factor of 2,
reaching the equilibrium conditions, and remaining constant
for the rest of the COSY cycle.

\begin{figure}[hbt]
 \leavevmode
 \unitlength 1.0cm
  \begin{picture}(12.2,14.0)
    \put(3.0,0.0){
       \epsfig{figure=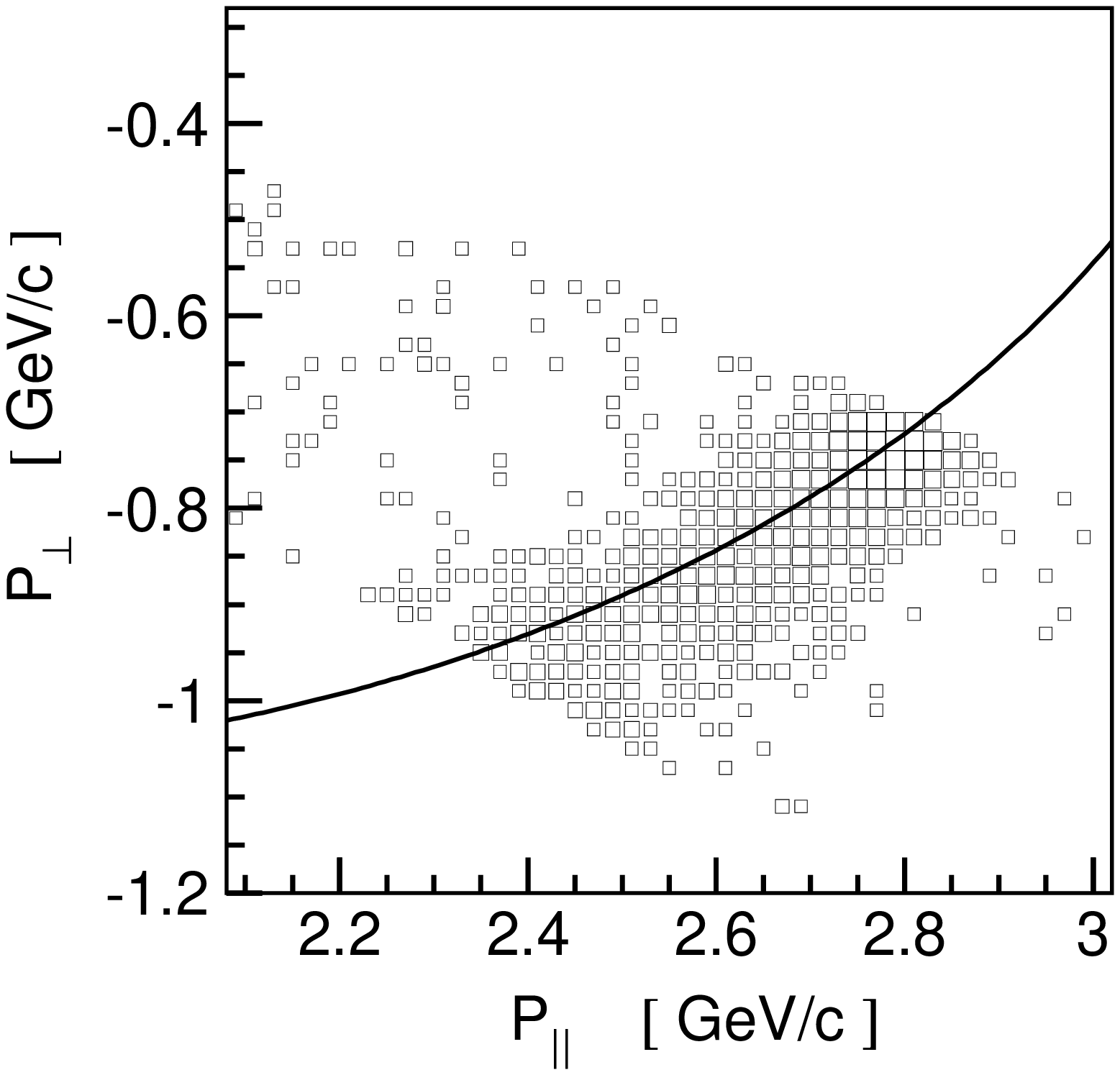,height=9.0cm,angle=0}
    }
       \put(10.7,1.5){
          { b)}
       }
    \put(3.0,7.0){
       \epsfig{figure=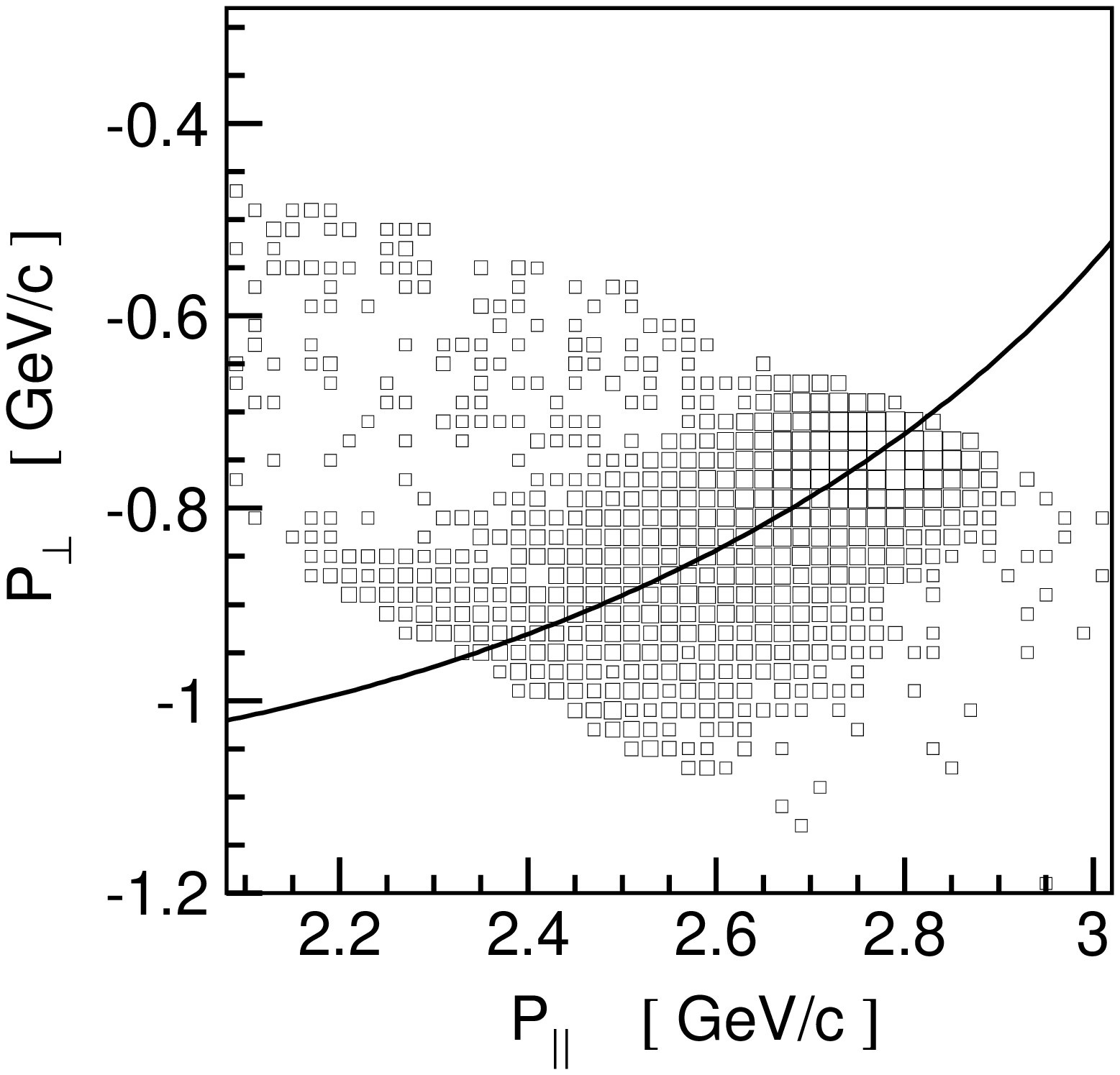,height=9.0cm,angle=0}
    }
       \put(10.7,8.5){
          { a)}
       }
  \end{picture}
\caption{  Perpendicular versus parallel momentum components with
           respect to the beam direction of particles registered 
           at a beam
           momentum of 3.227~GeV/c, as measured during the first minute~(a),
           and the last minute (b) of the 60 minutes long COSY cycle.
           The data at the first minute were analyzed 
           with $\Delta_{X}$~=~-0.15~cm,
           and the data of the 60's minute 
           with $\Delta_{X}$~=~-0.25~cm,
           see figure~\ref{sx_sy_cykl}c.
           The number of entries per bin is shown on a
           logarithmic scale. The solid line corresponds 
           to the momentum ellipse
           expected for protons scattered elastically 
           at a beam momentum of 3.227~GeV/c.
        }
 \label{trans_paral_cykl}
\end{figure}

 Figure~\ref{trans_paral_cykl} depicts the accuracy 
of the momentum reconstruction,
reflected in the spread of the data, which is mainly due
to the finite horizontal size of the beam and target overlap.
The upper and lower panel correspond to the first and the last minute of 
the measurement cycle, respectively. The data were analyzed correcting 
for the relative target and beam shifts~$\Delta_{X}$
as determined by the method discussed above.
The movement of the beam relative to the target during the cycle 
is quantified in Figure~\ref{sx_sy_cykl}c. The shift of the beam
denotes also changes of the average beam momentum,
due to the nonzero dispersion at the target position.
The beam reaches stability after some minutes.
This can be understood as the equilibration between the energy losses
when crossing $1.6\cdot 10^{6}$ times per second
through the $H_{2}$ cluster target, and the power of the longitudinal
stochastic cooling which can not only diminish the  
spread of the beam momentum but also shifts it as a whole.
\section{Conclusions}
We presented a method which allows the proton beam dimensions and
its position relative to the target to be controlled for internal beam 
experimental facilities. The technique is based on the measurements 
of the momentum distribution of elastically scattered protons and its 
comparison with the distributions simulated with different beam and 
target conditions.
We demonstrated the applicability of the method in controlling the proton 
beam during measurement when stochastic cooling was used at
the synchrotron COSY for the first time.

{\bf{Acknowledgements}}\\
One of the authors (P.M.) acknowledges hospitality and financial support 
from the Forschungszentrum J\"ulich.
We appreciate the careful reading of the manuscript by J.N. Tan.\\
This research project was supported in part by
the BMBF (06MS881I),
the Bilateral Cooperation between Germany and Poland
represented by the Internationales B\"{u}ro DLR 
for the BMBF (PL-N-108-95)
and by the  Polish State Committee for Scientific Research,
and by FFE grants (41266606 and 41266654)
from the Forschungszentrum J\"{u}lich.
\newpage
%
%
%
%
%
%
%
%

%
%

\newpage

\newpage
 
\newpage

\newpage

\newpage

\newpage

\newpage

\end{document}